\documentclass{emulateapj}

\newcommand{\alfven}{Alfv{\'{e}}nic}

\slugcomment{}

\shorttitle{The Origin of Type {\sc{I}} Spicule Oscillations}
\shortauthors{D.B. Jess et al.}

\begin{document}

\title{The Origin of Type {\sc{I}} Spicule Oscillations}

\author{D. B. Jess}
\affil{Astrophysics Research Centre, School of Mathematics and Physics, 
Queen's University Belfast, Belfast, BT7~1NN, Northern Ireland, U.K.}
\email{d.jess@qub.ac.uk}

\author{D. J. Pascoe}
\affil{School of Mathematics and Statistics, University of St Andrews, St Andrews, 
Scotland, KY16 9SS, UK}

\author{D. J. Christian}
\affil{Department of Physics and Astronomy, California State University Northridge, 
Northridge, CA 91330, U.S.A.}

\and

\author{M. Mathioudakis, P. H. Keys, F. P. Keenan}
\affil{Astrophysics Research Centre, School of Mathematics and Physics, 
Queen's University Belfast, Belfast, BT7~1NN, Northern Ireland, U.K.}

\begin{abstract}
We use images of high spatial and temporal resolution, obtained with the Rapid 
Oscillations in the Solar Atmosphere instrument at the Dunn Solar Telescope, to 
reveal how the generation of transverse waves in Type~{\sc{I}} spicules is a 
direct result of longitudinal oscillations occurring in the photosphere. Here we show how 
pressure oscillations, with periodicities in the range 130 -- 440~s, manifest in 
small-scale photospheric magnetic bright points, and generate kink 
waves in the Sun's outer atmosphere with transverse velocities approaching the 
local sound speed. Through comparison of our observations 
with advanced two-dimensional magneto-hydrodynamic simulations, we provide 
evidence for how magneto-acoustic oscillations, generated at the 
solar surface, funnel upwards along Type~{\sc{I}} spicule structures, before undergoing 
longitudinal-to-transverse mode conversion 
into  waves at twice the initial driving frequency. The resulting kink modes are 
visible in chromospheric plasma, with periodicities of 65 -- 220~s, and amplitudes often 
exceeding 400~km. A sausage mode oscillation also arises as a consequence of the 
photospheric driver, which is visible in both simulated and observational time series. 
We conclude that the mode conversion and period modification is a direct consequence 
of the 90 degree phase shift encompassing opposite sides of the photospheric driver. 
The chromospheric energy flux of these waves are estimated to be 
$\approx$3$\times10^{5}$~W{\,}m$^{-2}$, which indicates that 
they are sufficiently energetic to accelerate the solar wind and heat the localized 
corona to its multi-million degree temperatures.
\end{abstract}

\keywords{methods: numerical --- magnetohydrodynamics (MHD) --- 
Sun: atmosphere --- Sun: chromosphere --- Sun: oscillations --- Sun: photosphere}

\section{Introduction}
\label{intro}

\begin{figure*}
\centering
\includegraphics[width=14cm]{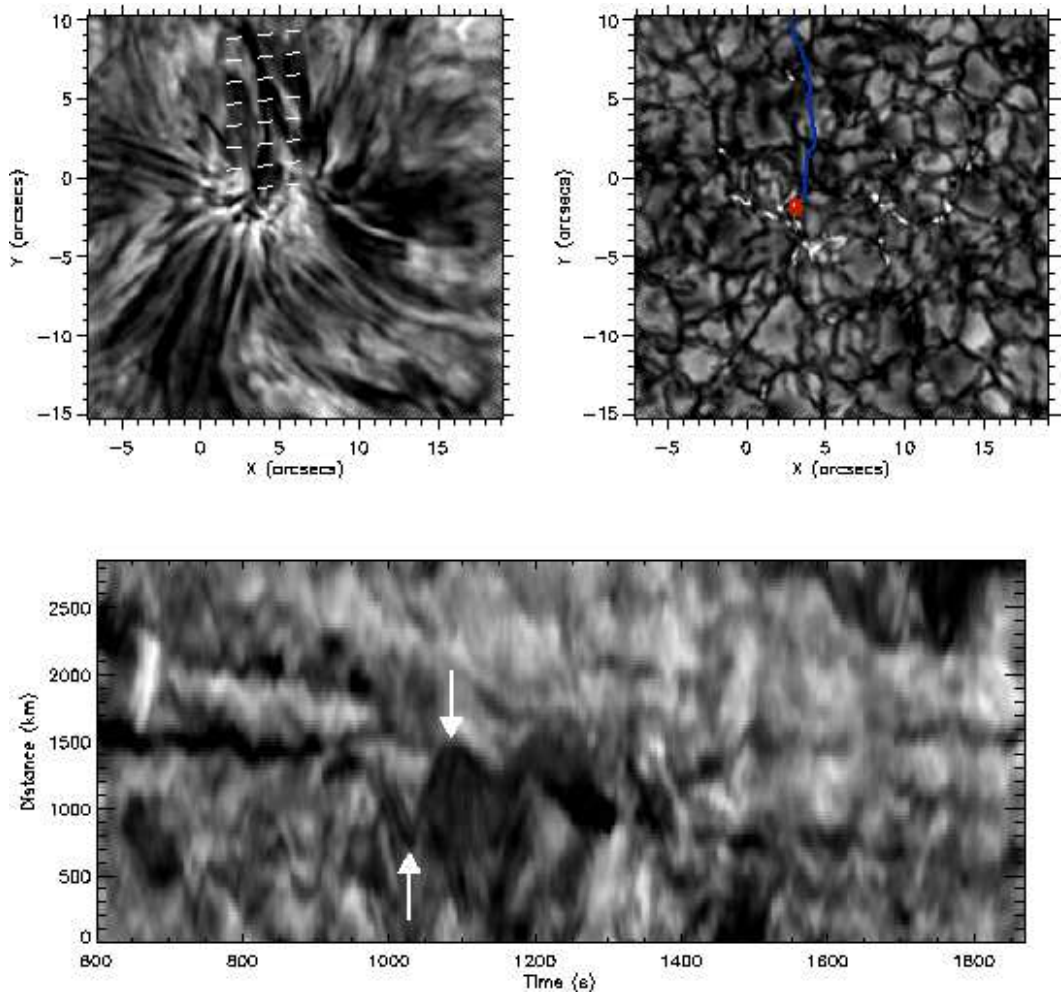}
\caption[]{Simultaneous images of the H$\alpha$ core (chromosphere; upper left) 
and G-band (photosphere; upper right), acquired at 14:06:21~UT on 28 May 2009. 
Dashed white lines in the H$\alpha$ core image highlight the spatial positions where 
time-distance cuts are made. The blue trace in the G-band image denotes the position of a
chromospheric spicule (visible in the left panel as a dark, straw-like structure), 
where one end is anchored into the photosphere above a MBP. 
A red contour indicates the location where a high concentration of longitudinal
oscillatory power is present. The lower panel is a sample H$\alpha$ time-distance cut, 
obtained 4000~km (5.5 arcsec) from the underlying MBP, revealing an abundance of periodic 
transverse motions in the solar chromosphere (see also Supplementary Movies 1 \& 2). 
White arrows highlight a trough and a peak of a typical transverse oscillation. The scale
is in heliocentric coordinates, where 1 arcsec approximately equals 725 km.}
\label{fig1}
\end{figure*}


The origin of the energy required to power the solar wind, and heat the Sun's corona 
to its multi-million degree temperatures, remains an elusive mystery in physics. 
One promising mechanism involves the creation of {\alfven} waves near the solar surface, 
which can penetrate upwards through the Sun's atmosphere with minimal reflection 
or energy loss \citep[][]{Erd07}. To date, there has been great controversy surrounding the interpretation 
of periodic transverse motions as Alfv{\'{e}}n waves. 
Magneto-hydrodynamic (MHD) waves with a wave number, $m = 1$, can be considered ``kink'' oscillations 
when structured by a cylindrical waveguide. The associated periodic transverse motions imply a degree 
of plasma non-uniformity, resulting in intrinsic MHD waves exhibiting mixed 
(i.e. slow, fast, Alfv{\'{e}}n) properties depending on the 
local plasma parameters. As a result, the term ``{\alfven}'' was introduced to describe waves 
which are predominantly characterised by the signatures displayed by 
pure Alfv{\'{e}}n waves. Accordingly, transverse kink oscillations may often be 
considered {\alfven} in nature providing they are incompressible, exhibit no intensity 
fluctuations along the structure, and display periodic displacements with the magnetic tension as the restoring 
force \citep[][]{Goo09}. 
Thus, by definition, {\alfven} waves require strong magnetic field concentrations, and/or a steep density 
gradient with the external plasma, to 
act as waveguides \citep[][]{Van08}, allowing energy to be directly channelled through the solar 
atmosphere. In the Sun's atmosphere, magnetic field lines clump together into tight bundles, 
forming flux tubes. The ubiquitous nature of {\alfven} waves in magnetic flux tubes has been 
demonstrated in a range of chromospheric and coronal plasmas 
\citep[][]{DeP07b, Tom07, Ban09, Jes09}. 

Spicules are dynamic, straw-like magnetic structures found in the solar chromosphere,
and can be divided into two distinct classes \citep[][]{Zaq09}. We will focus on Type~{\sc{I}} spicules, which
are ubiquitous throughout the solar atmosphere, and longer lived ($\approx$10~min) than their
jet-like Type~{\sc{II}} counterparts \citep[][]{DeP07a}. The narrow shape of all spicule structures ($<$600~km),
coupled with their rapidly evolving characteristics, places an utmost requirement for
both high-spatial and -temporal resolution instrumentation. Recently, \cite{DeP07b} and \cite{He09} 
observed Type~I spicules, which displayed periodic transverse motions with periods ranging from
50 -- 500~s, coupled with velocity amplitudes exceeding 25~km{\,}s$^{-1}$. Unfortunately, such
image sequences of spicules have been tied to limb observations. While observations at
the solar limb allow spicule height information to be easily retrieved, they do not allow
simultaneous measurements of the Sun's photosphere to be studied due to inherent 
line-of-sight effects. However, \cite{DeP11} and \cite{McI11} have recently shown how 
spicule dynamics make them ideal candidates in the quest for discovering efficient energy 
propagation into higher layers of the Sun's atmosphere.

\begin{table*}
\begin{center}
\caption{Spicules observed in both Ca~{\sc{ii}}~K and H$\alpha$ datasets. 
\label{table1}}
\begin{tabular}{lccccc}
~&~&~&~&~ \\
~		& Spicule  		& Kink	 	& Kink	 		& Peak Kink	 		& ~				\\
Spicule	& Length 			& Period 	 	& Amplitude		& Velocity  			& MBP Anchor 	\\
Number	& (km)			& (s)			& (km) 			& (km{\,}s$^{-1}$)		& Location		\\
~&~&~&~&~ \\
\colrule
1		& 4700	& 220	& 670	& 19.2 	& ($8''$, $-1''$) 	\\
2 		& 5200	& 139 	& 630	& 28.3 	& ($0''$, $-2''$)  	\\
3 		& 3900	& 65		& 160	& 14.8	& ($1''$, $-2''$)  	\\
4 		& 4100 	& 158 	& 410 	& 16.2	& ($3''$, $-4''$) 	\\
5$^{a}$	& 6200	& 129	& 380	& 18.5	& ($3''$, $-2''$) 	\\
6		& 4800	& 105	& 200	& 11.8	& ($8''$, $-4''$)	\\
7$^{b}$	& 5100	& 171 	& 190	& 7.2	& ($8''$, $-5''$)	\\
~&~&~&~&~ \\
\end{tabular}
\footnotesize \\
$^{a}$: Spicule used for display purposes in Figures~\ref{fig1}--\ref{fig3}. \\
$^{b}$: Spicule used for display purposes in Figure~\ref{fig4}. \\
\end{center}
\end{table*}

While transverse kink motion has been shown to be abundant in spicule observations, 
the underlying cause of the periodic motions has remained speculative. 
Overshooting of convective motions in the photosphere, granular buffeting, 
rebound shocks, and global p-mode oscillations have all been suggested as 
candidates for the creation of spicule oscillations \citep[][]{Rob79, Ste88, Vra08}. 
It is known that oscillations in the lower solar atmosphere are prone to aspects of 
mode coupling, whereby one particular magneto-acoustic wave mode 
\citep[normally classified as `fast' or `slow';][]{Nak05} may couple, and hence 
transfer wave energy, into another mode when atmospheric conditions are suitable 
\citep[][]{Ulm91, Kal97}. The rapid drop in density above the solar surface, 
coupled with the close proximity to oscillatory drivers, makes the photosphere and 
chromosphere an ideal location for the occurrence of efficient mode coupling. 
Indeed, \cite{McA03} uncovered upward-propagating transverse oscillations, 
generated in large ($>$2000~km diameter) photospheric bright points, which 
coupled into longitudinal waves at chromospheric heights. 
Often, the coupled oscillations demonstrate power at frequencies equal to twice the 
original, as a result of non-linear interactions \citep[][]{Ulm91}. Contrarily, it has 
been demonstrated, both analytically and theoretically, that when the magnetic 
pressure is approximately equal to the gas pressure (plasma $\beta=1$) in a solar flux tube, 
longitudinal to transverse mode coupling may also occur \citep[][]{DeM04}. 
To date, however, this form of ``reversed mode coupling'' has not been verified 
observationally. In this Letter, we undertake a multi-wavelength approach, allowing 
Type~{\sc{I}} spicules to be traced back to their anchor points on the solar surface. We provide 
conclusive evidence on how these ubiquitous oscillations manifest, and suggest 
what role they may take in the localized heating of the solar corona. 

\section{Observations}
\label{obs}
The data presented here are part of an observing sequence obtained during 
13:46 -- 14:47~UT on 2009 May 28, with the Richard B. Dunn Solar Telescope 
(DST) at Sacramento Peak, New Mexico. The newly-commissioned 
Rapid Oscillations in the Solar Atmosphere \citep[ROSA;][]{Jes10b} 
six-camera system was employed to image a $69.3''\times69.1''$ region 
positioned at solar disk center. A spatial sampling of $0.069''$ per pixel was 
used for the ROSA cameras, to match the telescope's diffraction-limited 
resolution in the blue continuum to that of the CCD. This results in images obtained at longer 
wavelengths being slightly oversampled. However, this was deemed 
desirable to keep the dimensions of the field-of-view the same for all 
ROSA cameras. 

During the observations presented here, 
high-order adaptive optics \citep[][]{Rim04} 
were used to correct wavefront deformations in real-time. The acquired images were 
further improved through speckle reconstruction algorithms \citep[][]{Wog08}, 
utilizing $16 \rightarrow 1$ restorations, providing reconstructed cadences of 
0.528~s (G-band) and 4.224~s (Ca~{\sc{ii}}~K and H$\alpha$). More details of our 
instrumentation setup can be found in \cite{Jes10a}. Atmospheric seeing conditions remained 
excellent throughout the time series. However, to insure accurate coalignment in all 
bandpasses, broadband time series were Fourier co-registered and de-stretched using a 
$40\times40$ grid, equating to a $\approx$1.7$''$ separation between spatial samples 
\citep[][]{Jes07,Jes08}. Narrowband images were corrected 
by applying destretching vectors established from simultaneous broadband reference 
images \citep[][]{Rea08}. 

\section{Analysis and Discussion}
\label{analy}
The ROSA field-of-view shows a range of features, including exploding granules and a
multitude of magnetic bright points (MBPs; Fig.~\ref{fig1}). A large conglomeration of 
MBPs were located at heliocentric coordinates (4 arcsec, $-3$ arcsec), or S01W00 in the 
solar north-south-east-west coordinate system, providing an opportunity to examine 
these kiloGauss structures without the line-of-sight spatial offsets normally 
associated with near-limb observations. Examination of
H$\alpha$ images reveals a wealth of chromospheric jets and spicules, anchored into the solar
surface at the locations of MBPs. These data reveal how un-anchored ends of magnetic 
flux tubes are observed to sway periodically over many complete cycles
(Supplementary Movies 1 \& 2). To quantify the oscillatory parameters, a series of one-dimensional
slits, each separated by 1 arcsec (725 km), were placed perpendicular to
the Type~{\sc{I}} spicules along their entire length (upper panels of Fig.~\ref{fig1}). 
The resulting time-distance cuts reveal numerous
periodic transversal displacements of these chromospheric structures as a function of
time. The sinusoid fitting algorithms of \cite{Bal11} were used to extract periodic signatures from
the transversal displacements present in the time-distance plots. We concentrate on the
seven most prominent Type~{\sc{I}} spicules, observed in both Ca~{\sc{ii}}~K and H$\alpha$ time-series, which
can be accurately traced back to their photospheric counterparts. The lower panel of 
Figure~\ref{fig1} displays a sample time-distance cut, revealing how these spicules
demonstrate transverse kink oscillations above their corresponding MBP in the period range
65 -- 220~s, with displacement amplitudes reaching as high as 670~km. Velocity amplitudes often 
exceed 15~km{\,}s$^{-1}$, indicating these dynamic motions are close to the chromospheric
sound speed \citep[][]{Fos05}. Table~\ref{table1} displays key characteristics of the spicules 
under investigation.

\begin{figure*}
\centering
\includegraphics[width=15cm]{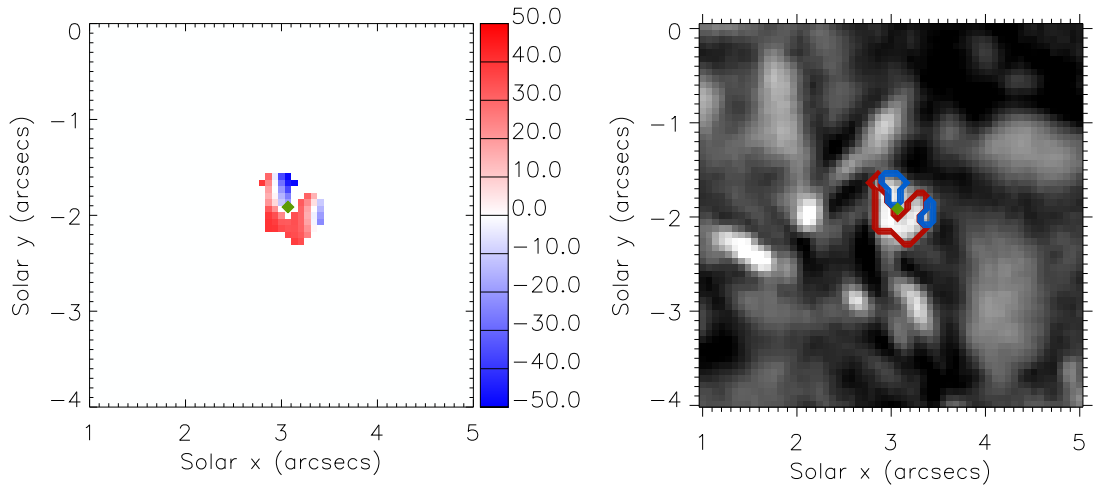}
\caption[]{Zoom-in of a MBP group acting as the photospheric anchor point of a 
chromospheric spicule. Maximum longitudinal oscillatory power is found at the center of the 
MBP (green pixel; left panel), equating to power in excess of $7\times10^{5}$ times the 
quiescent background. Phase shifts (in degrees), relative to this local maximum, are 
displayed using the red/blue color table, with maximum values in this example of $+40$ 
and $-49$ degrees. The right panel displays the phase information contoured on top of 
the simultaneous G-band intensity image, revealing how oscillatory power is closely 
confined within the MBP perimeter.}
\label{fig2}
\end{figure*}


No periodic transverse motions in either the MBP edge, or bright central peak, were detected in 
simultaneous photospheric G-band images, implying that
either spicule motion is generated independent of processes manifesting on the solar
surface, or that a form of efficient mode conversion between the two layers
(photosphere and chromosphere) is occurring. To investigate the 
latter, G-band intensities were averaged over the entire 
MBP structure under investigation, with fast Fourier transforms and wavelet analysis 
routines subsequently applied on the resulting intensity fluctuations. 
We find significant oscillatory power (exceeding 500 
times the quiescent background) in the averaged intensity of these G-band features, which 
typically occupy $\sim$50~pixels, or 130{\,}000~km$^{2}$. 
Furthermore, the most powerful intensity oscillations found in each photospheric MBP are at twice the 
period (130 -- 440~s) of the corresponding chromospheric transverse oscillation, indicating the 
presence of frequency doubling. The dominant period remains  
constant over each MBP surface, indicating the presence of powerful, yet coherent 
drivers. We further utilize our high-resolution observations to deduce additional parameters 
related to the longitudinal photospheric drivers, and find that the peak longitudinal oscillatory power 
is located at the center of each photospheric MBP, with a magnitude exceeding $7\times10^{5}$ 
times the background average (right panel of Fig.~\ref{fig2}). Power decreases 
radially away from the center of the photospheric 
MBP, eventually dropping to zero at the edge of the structure. We compute the spatial distributions 
of oscillatory phase for the photospheric MBPs relative to this position of peak power. 
Figure~\ref{fig2} reveals how deviations in phase angle, from the positions of peak-power, are 
within the range of $+40$ to $-49$~degrees, providing absolute phase shifts of 
$\approx$90~degrees across each entire MBP structure. These phase shifts remain constant throughout the 
duration of periodic behaviour, with oscillatory power eventually ceasing as the MBP 
fragments. In addition, the detected photospheric MBP periodicities (130 -- 440~s) overlap with the 
solar p-mode spectrum \citep[][]{Lei62, Ulr70}, suggesting these global modes 
provide the underlying driving force required to generate spicule oscillations.

\begin{figure}
\centering
\includegraphics[width=\columnwidth]{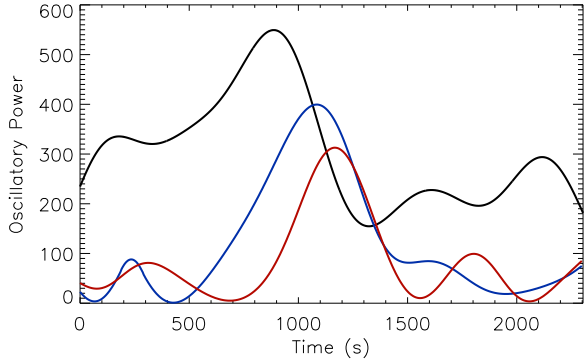}
\caption[]{Oscillatory power of an
isolated MBP group in G-band (black), and associated Ca~{\sc{ii}}~K (blue), and H$\alpha$ (red)
spicule time series. The x-axis is time from the start of the observing sequence, while
the y-axis is power averaged over the entire MBP/spicule structure, and normalized to
the background quiescent Sun. G-band power is for oscillations in intensity at a period
of 258~s, while the Ca~{\sc{ii}}~K and H$\alpha$ power is derived from the periodic transverse
displacements of the spicule with a periodicity of 129~s. It is clear the G-band power
leads the Ca~{\sc{ii}}~K and H$\alpha$ time-series, indicating the presence of an upward propagating
magneto-acoustic wave, which couples into an transverse kink mode near the chromospheric
boundary.}
\label{fig3}
\end{figure}

To investigate the temporal variations in wave behaviour, we calculate the oscillatory power 
at each time step, and compare with simultaneous measurements of the same 
spicule observed in different wavelengths. 
Our analysis shows that the power of lower chromospheric (Ca~{\sc{ii}}~K) transverse 
oscillations peaks after the photospheric (G-band) longitudinal periodicities, yet before the 
upper chromospheric (H$\alpha$) oscillations. Figure~\ref{fig3} reveals how a 
discernible time lag exists between three distinct layers of the lower solar atmosphere. 
This implies that the observed chromospheric spicule 
oscillations are a result of upwardly propagating magneto-acoustic wave modes, 
generated at the solar surface, which couple into transverse kink waves in a region encompassing 
the photospheric/chromospheric boundary. An important trend to consider is how the power 
of detected oscillatory behaviour varies as a function of
atmospheric height. Power spectra for the G-band intensity, and Ca~{\sc{ii}}~K and 
H$\alpha$ transverse oscillations, 
were normalised to their respective quiescent backgrounds before comparison. Figure~\ref{fig3} 
shows an appreciable decrease in oscillatory power as a function of atmospheric height. 
This has important consequences, as it suggests the presence of physical
damping, and therefore conversion into heat, of the observed wave modes.
Further examination of each MBP reveals multiple power peaks (also visible in Fig.~\ref{fig3}), 
suggesting the creation, and
subsequent transverse behaviour of solar spicules, is quasi-periodic in nature. 
This is the first observational evidence of longitudinal-to-transverse 
mode coupling occurring in the Sun's atmosphere.

\begin{figure}
\centering
\includegraphics[width=\columnwidth]{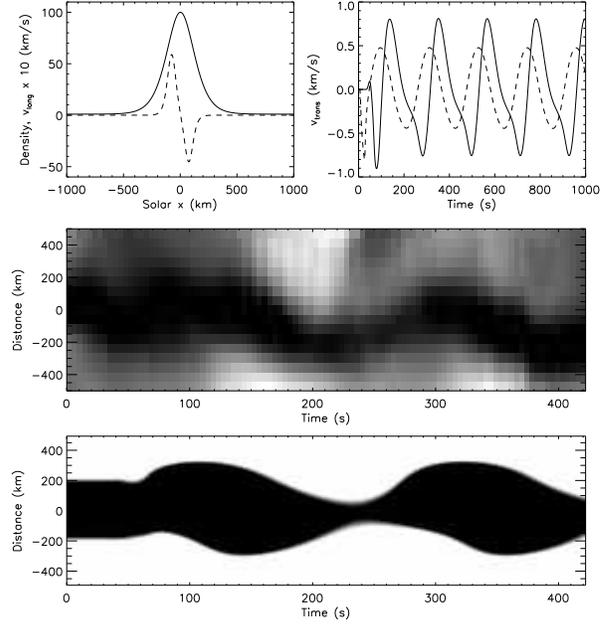}
\caption[]{{\it{Upper left}}: The spicule density profile (solid line) normalised to the background value, while the 
dashed line shows a snapshot of the longitudinal velocity amplitude, 130~s into the simulation, and 
scaled for clarity. {\it{Upper right}}: The transverse velocity signal, measured 
at the spicule axis, as a function of time for two heights. The dashed and solid lines represent the signal 
obtained at heights of 500~km and 2000~km above the photosphere, respectively. A periodic 
transverse velocity signal, which increases in amplitude away from the photosphere, reveals the 
progression of longitudinal-to-transverse mode coupling. Frequency doubling, represented by a change from a
purely sinusoidal waveform, is observed in the signal from a height of 2000~km. 
{\it{Middle}}: A time-distance cut of a chromospheric 
spicule observed in H$\alpha$. {\it{Bottom}}: A time-distance cut of the simulated spicule density, at a height of 
2000~km above the photospheric boundary. There is a remarkable degree of similarity between the two lower 
panels, with both kink (transverse displacement of spicule axis) and sausage (periodic compression and 
rarefaction) modes visible. Supplementary Movies 3 \& 4 display time-lapse image sequences of the 
simulations presented here.}
\label{fig4}
\end{figure}

To further investigate the propagation and coupling of these wave phenomena, we utilized 
advanced two-dimensional magneto-hydrodynamic (MHD) simulations. 
These simulations were performed using the Lare2D code \citep[][]{Arb01}, where a 
spicule was modelled as a density enhancement, embedded in a straight, uniform magnetic field. The 
spicule density has a symmetric Epstein profile \citep[][]{Nak95} in the transverse direction, with 
a peak density 100 times the background value, and a diameter of approximately 
600~km (upper-left panel of Fig~\ref{fig4}). The magnetic field strength is chosen to give an external 
Alfv{\'{e}}n speed of 362~km{\,}s$^{-1}$, and an Alfv{\'{e}}n speed of 
36.2~km{\,}s$^{-1}$ at the centre of the spicule. A 
temperature profile is chosen to give a plasma $\beta=1$, where efficient mode coupling is 
expected to occur \citep[][]{DeM04}. The simulation has a resolution of $600\times2000$ 
grid points, providing a numerical domain size of $4000\times20000$~km$^{2}$. For the 
purposes of this investigation, only effects up to a few thousand~km (i.e. a region incorporating 
the photosphere and chromosphere) will be considered.

An artificial driver is applied at the lower boundary in the form of a longitudinal velocity 
component. This is consistent with the observational interpretation that the driver 
arises from a global p-mode oscillation. The coherent driver is applied across the diameter 
of the spicule, with a periodicity of 215~s, and a maximum amplitude of 12.5~km{\,}s$^{-1}$. 
However, to remain consistent with parameters derived from the observational time series, 
one side of the simulated MBP has a phase difference of 90~degrees with respect to the other 
(upper-left panel of Fig.~\ref{fig4}). 
As the driver is longitudinal and compressive, a 90~degree phase 
shift creates transverse (i.e. across the spicule) gradients in pressure which displace the 
spicule axis, generating the kink mode, in addition to producing periodic 
compressions and rarefactions (sausage mode). These periodic pressure differences, 
across the body of the spicule, also induce a frequency doubling of the coupled 
transverse wave, visible in the upper-right panel of Figure~\ref{fig4} by a change from a purely 
sinusoidal waveform.
Both transverse kink and sausage modes are readily apparent in the lower panels of Figure~\ref{fig4}, where 
time-distance cuts of observational, and simulated data, are displayed (see also Supplementary 
Movies 3 \& 4). In our simulations, the magnitude of the excited kink mode is smaller than the 
amplitude of the input longitudinal driver. This is a consequence of the uniform magnetic 
field strength used. In the real solar atmosphere, the magnetic field strength 
decreases rapidly with height, which is a likely contributor to the more prominent increase in 
kink amplitude seen in current ground- and space-based observations of chromospheric spicules 
\citep[e.g.][]{He09}. Nevertheless, the upper-right panel of Figure~\ref{fig4} indicates that the 
longitudinal-to-transverse mode-coupling mechanism present in our MHD simulations does 
promote a progressive increases in kink amplitude as a function of atmospheric height.

By combining observed 
photospheric periodicities and phase distributions, in conjunction with realistic MBP 
structuring, our simulations reveal efficient, and ubiquitous, mode conversion of 
longitudinal oscillations into their transverse kink counterparts, at twice the driven frequency. 
The simulations display wave properties consistent with our 
observations, and demonstrate how transverse spicule oscillations, with significant 
amplitudes, readily exist in the solar atmosphere, and therefore have important 
consequences for energy transportation into the solar corona. 
We estimate the energy flux, E, of the observed chromospheric waves using 
\citep[][]{DeP07b},
\begin{equation}
\mathrm{E} = \rho v^{2} v_{A} \ ,
\end{equation}
where $\rho$ is the mass density of the flux tube, $v$ is the observed velocity amplitude and $v_{A}$ 
is the corresponding Alfv{\'{e}}n speed, defined as $v_{A} = B/\sqrt\mu_{0}\rho$, with $\mu_{0}$ 
the magnetic permeability. For a mass density of $\rho\approx1.3\times10^{-8}$~kg{\,}m$^{-3}$, 
and a magnetic field strength $B\approx12$~G, derived from a bright network chromospheric 
model \citep[][]{Ver81}, coupled with an 
observed velocity amplitude of $v\approx15$~km{\,}s$^{-1}$, the energy flux in the 
chromosphere is $E\approx3\times10^{5}$~W{\,}m$^{-2}$. It is estimated that 
approximately 2.2\% of the solar surface is covered by 
MBPs \citep[][]{San10}, and if each MBP is linked to a single corresponding chromospheric Type~{\sc{I}} spicule, it 
equates to an average global energy of 660~W{\,}m$^{-2}$. Current work suggests waves with an 
energy flux $\approx$100~W{\,}m$^{-2}$ are sufficiently vigorous to heat the localized corona and launch 
the solar wind when their energy is thermalized \citep[][]{Jes09}. 
Therefore, a transmission coefficient of $\approx$15\% through the thin transition region 
would provide sufficient energy to heat the 
entire corona. Regions on the solar surface containing highly magnetic structures, such 
as sunspots, pores and large MBP groups, should possess even higher mass densities 
and magnetic field strengths, in addition to a greater numbers of spicules. In this regime, 
the energy flux available to heat the corona will be significantly higher than the 
minimum value required to sustain localized heating.

\section{Concluding Remarks}
\label{conc}
We have utilized images of high spatial and temporal resolution, obtained with the Rapid 
Oscillations in the Solar Atmosphere (ROSA) instrument at the Dunn Solar Telescope, to 
reveal how the generation of transverse kink oscillations in Type~{\sc{I}} spicules is a 
direct result of mode conversion in the lower solar atmosphere. Through comparison of 
our observations with advanced two-dimensional magneto-hydrodynamic (MHD) simulations, 
we show how longitudinal pressure modes in photospheric magnetic bright points (MBPs), 
with periodicities in the range 130 -- 440~s, funnel upwards through the Sun's atmosphere, 
before converting into kink modes at twice the initial frequency, often with amplitudes exceeding 
400~km. We conclude that the mode conversion and period 
modification is a direct consequence of the 90 degree phase shift encompassing opposite 
sides of the photospheric driver. This is the first observational 
evidence of this mechanism occurring in the solar atmosphere. Energy flux estimates for these 
oscillations indicate that the waves are sufficiently energetic to accelerate the solar wind and heat 
the quiet corona to its multi-million degree temperatures. 

The naming of transverse oscillations 
observed in solar structures remains highly controversial, with definitions revolving around 
``Alfv{\'{e}}n'', ``{\alfven}'', and ``magneto-sonic kink'' terminology. 
While our observations demonstrate signatures consistent with 
previous studies on {\alfven} waves \citep[e.g.][]{DeP07b, McI11}, 
we have deliberately chosen to describe the observed periodic motions 
simply as transverse kink waves. This is currently the most unopposed description of 
such wave phenomena, and avoids the potential pitfalls of what is a 
rapidly developing area within solar physics.

\acknowledgments
DBJ wishes to thank STFC for the award of a Post-Doctoral Fellowship. 
DJC thanks CSUN for start-up funding related to this project. 
PHK is grateful to NIDEL for a PhD studentship. 
Solar Physics research at QUB is supported by STFC. The ROSA project is supported by EOARD.

{\it Facilities:} \facility{Dunn (ROSA)}.

\end{document}